\documentclass[twocolumn,preprintnumbers,amsmath,amssymb,floatfix,prd,superscriptaddress]{revtex4}

\usepackage{graphicx}
\usepackage{dcolumn}
\usepackage{bm}
\usepackage{hyperref}
\usepackage{natbib}
\usepackage{xcolor}
\definecolor{RED}{named}{red}
\usepackage{subcaption}
\usepackage{url}
\hypersetup{colorlinks,citecolor=cyan}
\hypersetup{
    colorlinks=true,
    linkcolor=cyan,
    filecolor=cyan,
    urlcolor=cyan,
    }

\begin{document}

\title{Charged galactic wormholes :  shadow imaging, accretion disks, and
charged-particle deflection}

\author{Md Khalid Hossain}
\email{mdkhalid.ac@gmail.com}
\affiliation{Department of Physics, Indian Institute of Technology Hyderabad, Kandi, Sangareddy, Telengana 502285, India}

\author{Humaun Kabir Laskar}
\email{humaunkabir302@gmail.com}
\affiliation{Department of Mathematics, Jadavpur University, Kolkata 700032, West Bengal, India}

\author{Farook Rahaman}
\email{rahaman@associates.iucaa.in}
\affiliation{Department of Mathematics, Jadavpur University, Kolkata 700032, West Bengal, India}

\author{Aritra Sanyal}
\email{aritrasanyal1@gmail.com}
\affiliation{Department of Mathematics, Jadavpur University, Kolkata 700032, West Bengal, India}

\date{\today}

\begin{abstract}
{We investigate the gravitational deflection of charged massive particles by a charged galactic wormhole supported by the Sofue dark matter density profile \cite{r57},
$\rho(r)=\rho_0 e^{-r/r_0},$
where $\rho_0$ and $r_0$ denote the central dark matter density and the characteristic scale radius, respectively. This phenomenological profile models the dark matter distribution in galactic halos and acts as the matter source sustaining the wormhole geometry. Extending our previous investigation of light rays and neutral massive particles in this spacetime \cite{b1}, we study the combined effects of gravitational and electromagnetic interactions on the motion of charged massive particles. The deflection angle is computed independently using the Rindler--Ishak method based on the Jacobi metric and the Gauss--Bonnet theorem, enabling a systematic comparison between the two approaches. We find that the predictions of the two methods become nearly indistinguishable in the relativistic regime, whereas the differences at lower velocities arise from the velocity-dependent correction terms proportional to $(1-v^2)$ in the trajectory equation. We further perform backward relativistic ray tracing of optically thin accretion flows around the charged galactic wormhole to investigate its shadow and photon-ring structure. Although the shadow remains circular because of the spacetime's spherical symmetry, its radius, photon-ring structure, and intensity distribution exhibit a nontrivial dependence on the wormhole charge, providing potential observational signatures of charged galactic wormholes. These results provide new insights into the interplay between gravitational and electromagnetic interactions in charged wormhole spacetimes and their observational manifestations.}
\end{abstract}

\maketitle

\textbf{Keywords:} Lensing; shadow; Charged wormhole; Deflection of charged particle

\section{Introduction}

In observational astrophysics, wormholes continue to be a promising area. One needs to interact with the geometrical structure of the wormhole in order to collect data through observational detection. Thus, scientists are trying to visualize wormholes in order to understand their geometrical features.
One of the first experimental validations of the general theory of relativity was gravitational lensing \cite{r1}. The way the sun deflects light was the first to be identified, followed by the way foreground galaxies lens quasars, the formation of huge arcs in galaxy clusters, and finally the galactic microlensing technique. In the realm of observatory astronomy, it is currently a common occurrence \cite{r2}.
The notion of wormholes and a non-simply interconnected space-time was first put forth by Fuller and Wheeler \cite{r3}. However, Einstein and Rosen had already studied the non-singular coordinate patches of the Schwarzschild and Reissner-Nordstrom solutions \cite{r4}.
Morris and Thorne later developed the idea of traversable wormholes that included exotic material \cite{r5}. Other scholars made additional changes \cite{r6,r7,r8}.
Although weak lensing from celestial objects has been extensively studied over the past century, lensing in the strong gravitational field regime is a relatively new topic of study that has attracted a lot of attention recently \cite{r9,r10,r11,r12,r13,r14,r15,r16}. It is now known that wormhole models can exist in modified gravity theories with normal matter, even though exotic matter is required for them to exist in General Relativity \cite{r17}-\cite{r46}.
It is very desirable to expand these research avenues, particularly in light of the preliminary results from the Event Horizon Telescope \cite{r47,r48}.
All of these tests have, in fact, provided a wealth of information about the BH signatures through the lensing process. However, another fascinating concept that hasn't received enough attention thus far is: Lensing from stellar-scale traversable wormholes (WH) is as intriguing as that from black holes (BHs). The literature has extensively studied WHs, which are areas of spacetime with throats linking two asymptotically flat regions. Light propagation is one of the many fascinating effects associated with WH spacetimes, and it has been analyzed in particular detail for the Morris--Thorne - (MT) WH spacetime \cite{r5}, whose static, spherically symmetric line element admits closed-form photon trajectories and is therefore especially well suited for an explicit, analytically tractable study of light bending. We also investigate the shadow and photon-ring properties of the charged galactic wormhole by carrying out relativistic ray-tracing of optically thin accretion flows. The resulting shadow morphology and intensity distribution exhibit a strong dependence on the wormhole charge, indicating potential observational imprints relevant to future high-resolution observations.
Building on the early work of Cramer et al.\ \cite{r52} on the problem of lensing by negative mass WHs, Safonova et al.\ \cite{r51} did a new study of lensing of WHs. Most recently, Tejeiro and Larranaga found that Morris-Thorne type WHs generally behave as convergent lenses \cite{r53}.
Gravitational lensing is a crucial tool in astrophysics and cosmology for a variety of tasks, such as determining the cosmological constant, mapping the distribution of dark matter, measuring the Hubble constant, verifying the existence of extrasolar planets, and comprehending the mass distribution in the vast structure of the universe \cite{r55}. Recently, in our previous paper \cite{r56}, we investigated the deflection of a heavy object around the Joshi-Malafarina-Narayan (JMN) naked singularity.

This paper is organized as follows: In Sec.~\ref{1}, we discuss charged galactic wormholes. In Sec.~\ref{2}, we examine the deflection of massive objects around the wormhole using the Rindler--Ishak method. Sec.~\ref{3} focuses on calculating the deflection angle using the Gauss--Bonnet method. In Sec.~\ref{4}, we study the photon ring and shadow of charged wormholes. Finally, Sec.~\ref{5} presents the conclusion.

\section{Overview of the Charged Galactic Wormhole} \label{1}

It has been proposed recently that wormholes could exist in galactic regions \cite{q1,q2}. A new dark matter density profile for certain galaxies, known as the exponential density profile, was introduced by Sofue \cite{r57}. This model depicts how dark matter is distributed within galaxies and, in certain instances, aligns more closely with observed galactic rotation curves. The exponential profile is similar to the density distributions of stellar disks in spiral galaxies, suggesting a potential link between the distributions of visible and dark matter. It is especially useful for examining the concentration of dark matter in the inner parts of galaxies, including the galactic center, where traditional models often struggle. Our proposal involves investigating the backreaction on the charged galactic wormhole spacetime, using Sofue density profile to derive exact solutions.
By departing from conventional mass-dominated models, the charged galactic wormhole metric we propose challenges traditional ways of thinking. With the inclusion of Sofue density profile, the model could forecast significant phenomena. The charged galactic wormhole solution employed in this work was obtained by solving the Einstein--Maxwell field equations with the Sofue density profile as the matter source. The resulting spacetime is static and spherically symmetric, where $Q$ denotes the electric charge of the wormhole and $b(r)$ is a radial metric function determined by the field equations. The proposed solution satisfies the flare-out condition at the wormhole throat, thereby fulfilling one of the fundamental geometric requirements for a traversable wormhole. Moreover, consistent with traversable wormhole solutions in general relativity, the null energy condition is violated in the vicinity of the throat, indicating the presence of exotic matter required to sustain the wormhole geometry. In addition, the stability of the wormhole was investigated through a linearized stability analysis. A comprehensive discussion of these properties, including the flare-out condition, the null energy condition, and the stability analysis, can be found in our previous work, and the corresponding line element is given by \cite{b1}

\begin{equation}\label{e1}
ds^{2}=-\left(1+\frac{Q^2}{r^2}\right)dt^2 + \left(1-\frac{b(r)}{r}+\frac{Q^2}{r^2}\right)^{-1}dr^2+r^2 d \Omega^2,
\end{equation}
where\\ $d\Omega^2=d\theta^2+\sin^2\theta\,d\phi^2$,\\
$b(r)=-8r_s\!\left(r^2+2rr_s+2r^{2}_{s}\right)e^{-\frac{r}{r_s}} \pi \rho_{s}+C_1$.

$C_1$ is constant, $r_s$ denotes the scale radius, $\rho_s$ the central density, and $Q$ the electric charge of the wormhole.

\section{Charged Massive Particle (CMP) deflection via Rindler-Ishak Method} \label{2}

For a spherically symmetric metric
\begin{equation}
ds^2=-Adt^2+Bdr^2+C(d\theta^2+\sin^2\theta\,d\phi^2),  
\end{equation}

\begin{widetext}
the trajectory equation becomes \cite{0}
\begin{equation}
\left({\frac{dU}{d\phi}}\right)^2=\frac{C^2U^4}{A B v^2b^2}\!\left[\left(1+\frac{\sqrt{1-v^2}}{m}qA_{t}\right)^2\!-A\!\left(1-v^2+\frac{b^2v^2}{C}\right)\right]
\equiv f(U). \label{geo}
\end{equation}
\end{widetext}

Here, $U=1/r$ denotes the reciprocal radial coordinate, which allows the trajectory equation of the charged massive particle to be written in a more convenient form. $A_\mu$ denotes the electromagnetic four-potential associated with the charged wormhole. Throughout this work, we assume a purely electrostatic field, $A_\mu dx^\mu=-Q\,dt/r$, so that the scalar potential is $A_t=-Q/r$, where $Q$ is the electric charge of the wormhole. This reduces to the null geodesic equation for $v\to 1$.

\begin{figure}[htbp]
\centering
\includegraphics[scale=0.19]{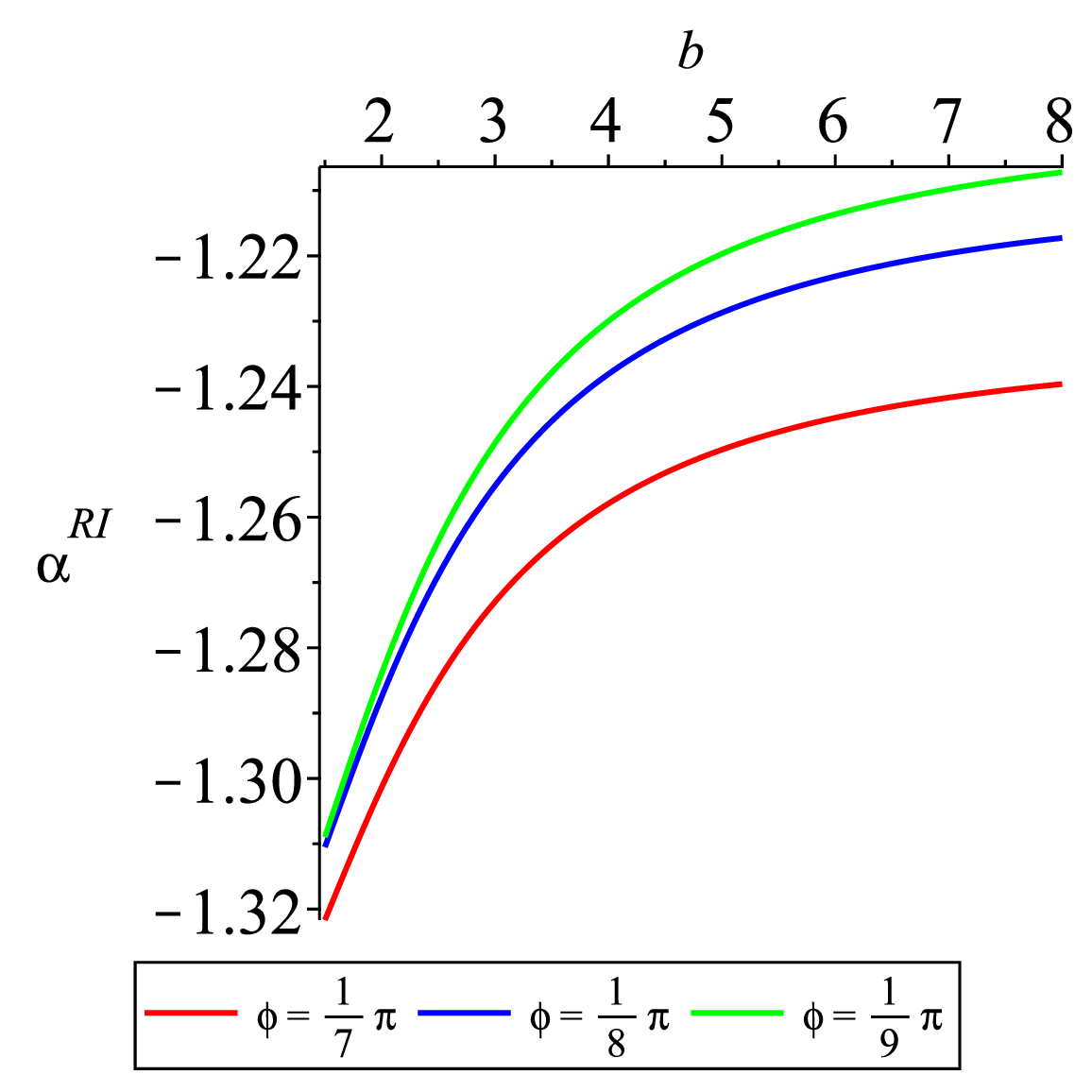}
\includegraphics[scale=0.19]{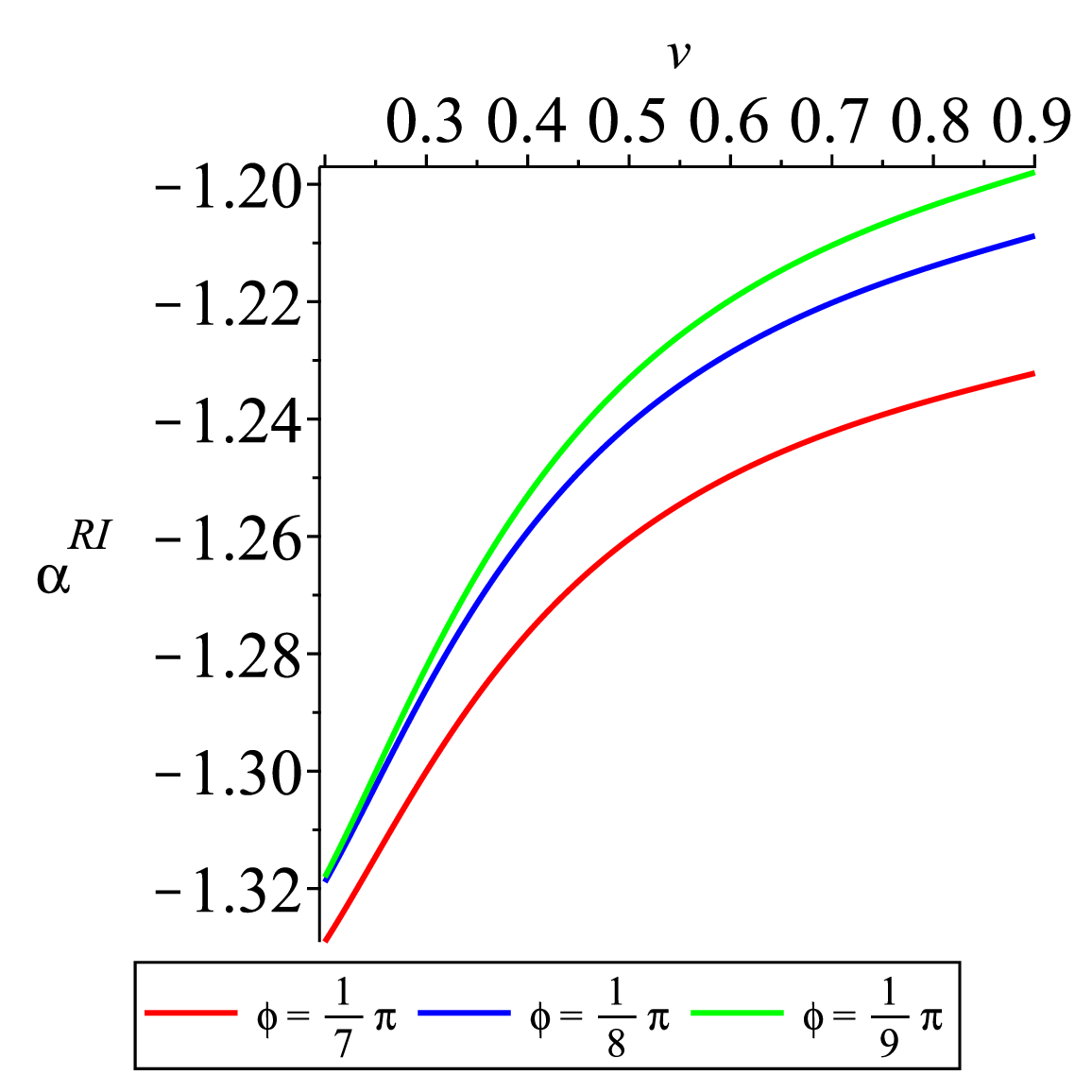}\\[-2pt]
\includegraphics[scale=0.19]{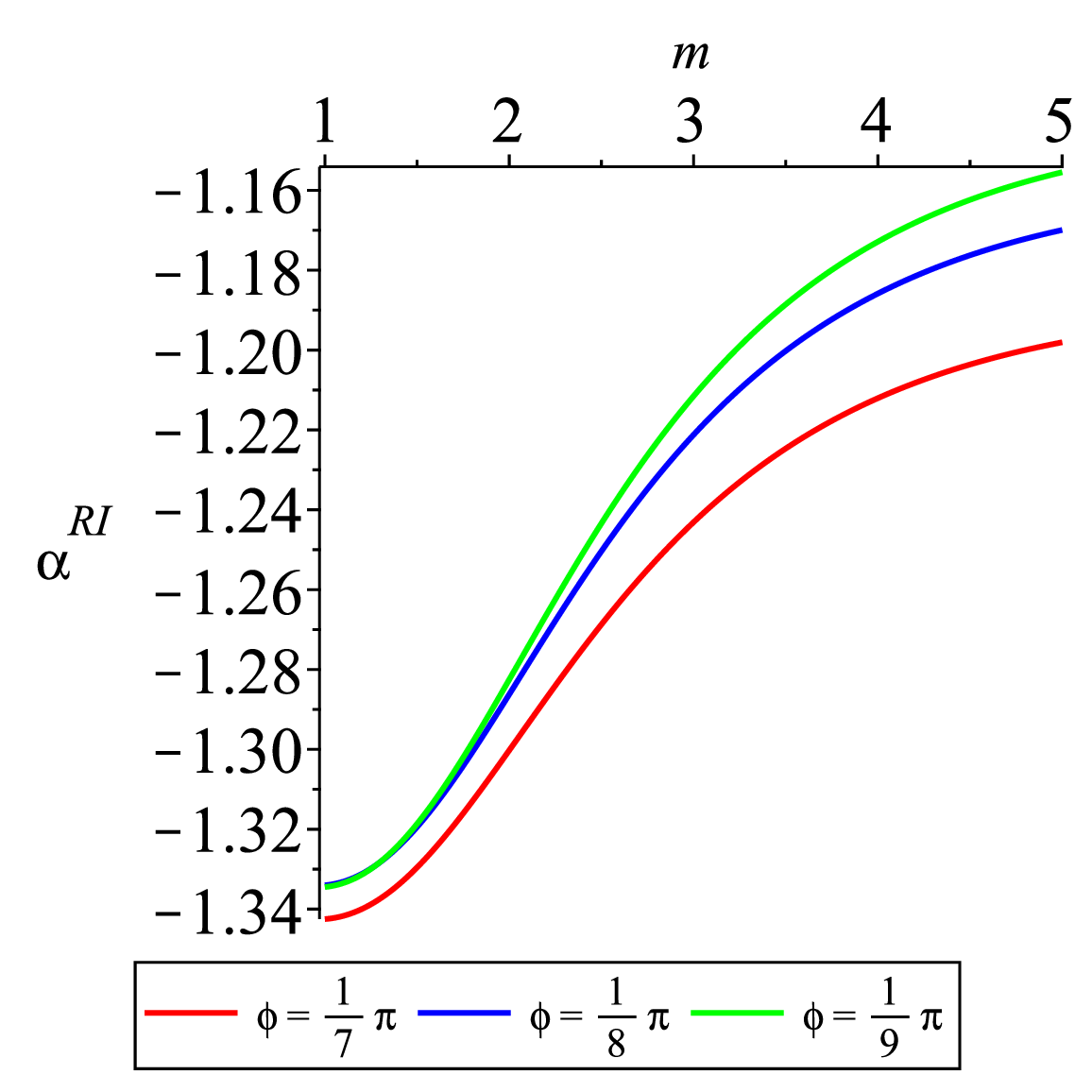}
\includegraphics[scale=0.19]{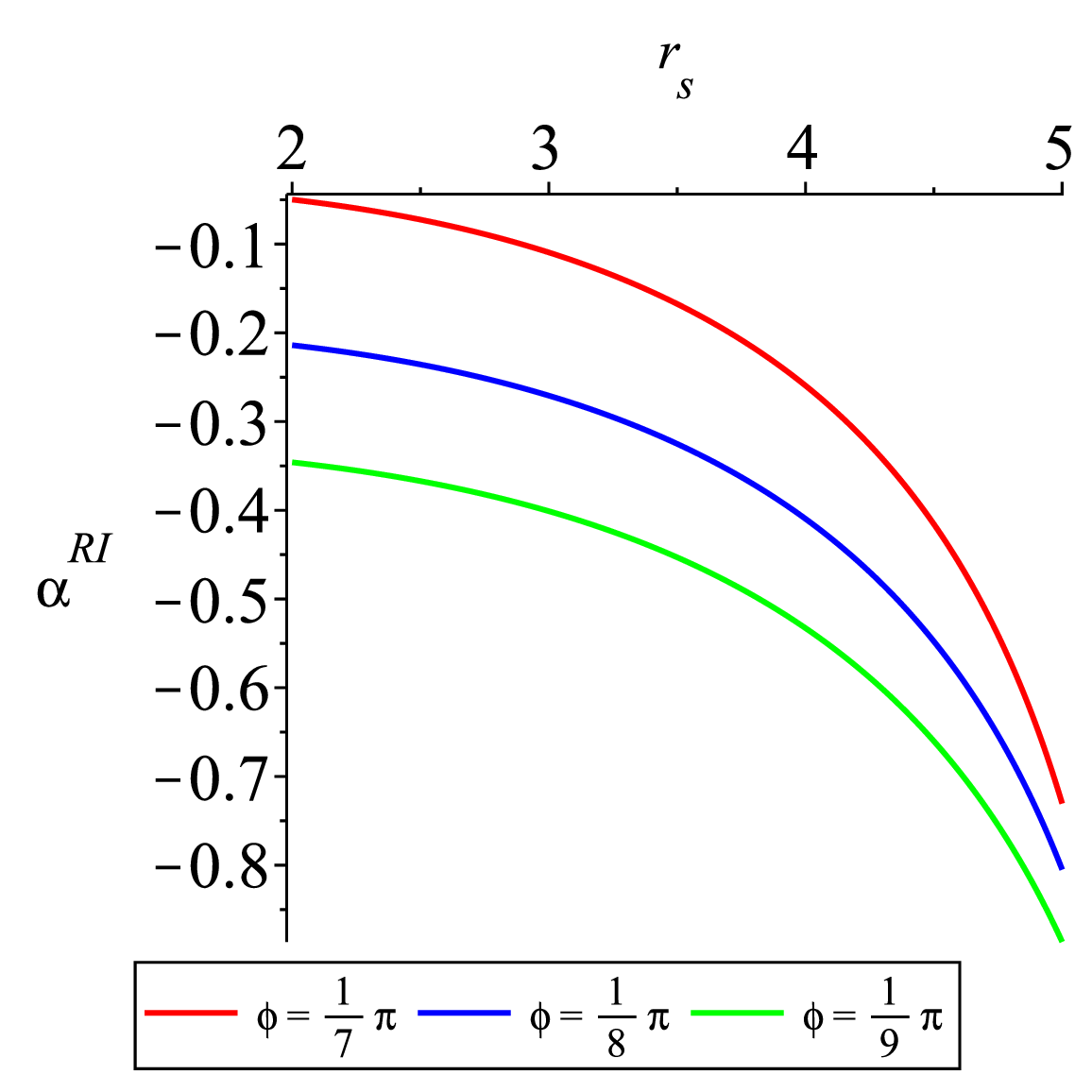}\\[-2pt]
\includegraphics[scale=0.19]{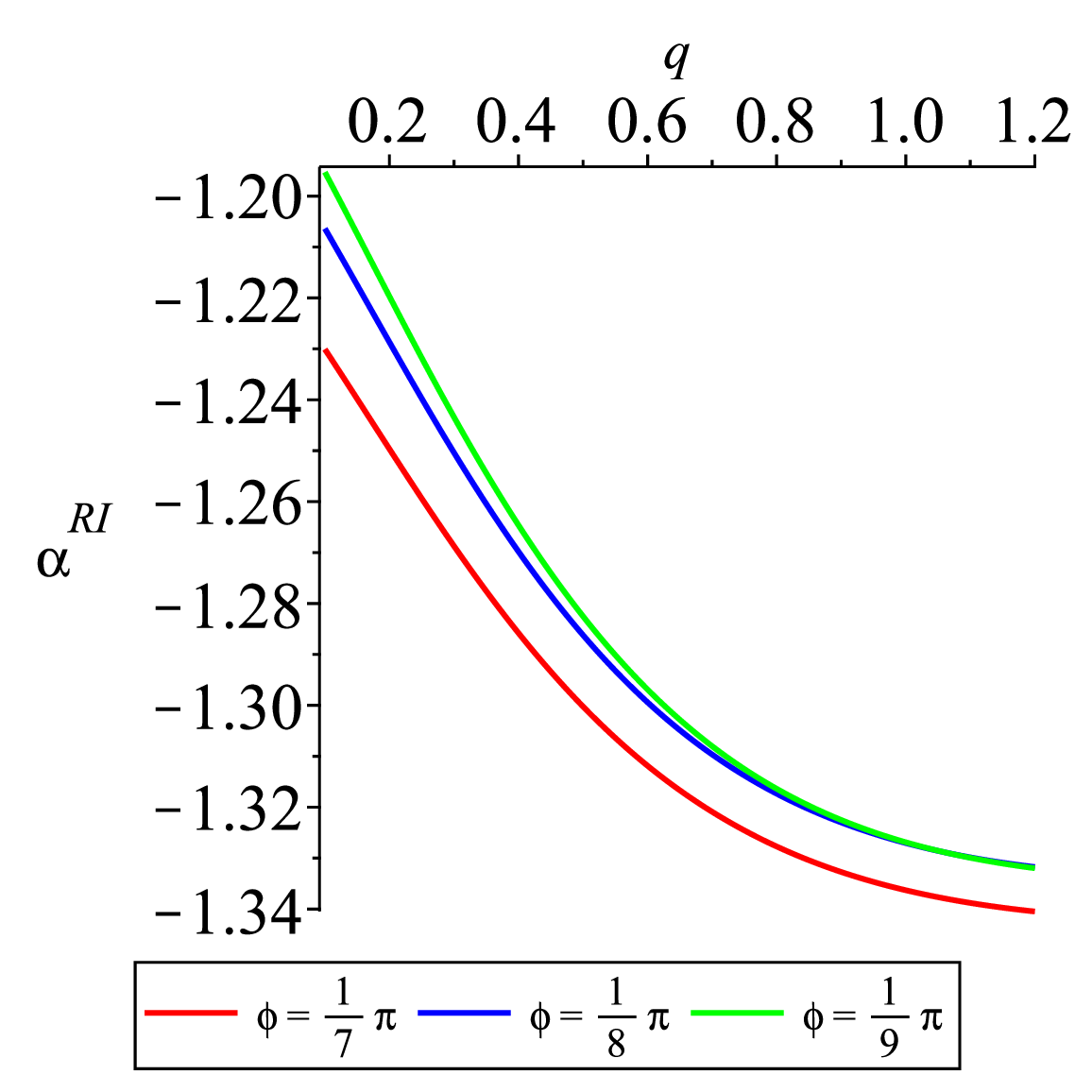}
\includegraphics[scale=0.19]{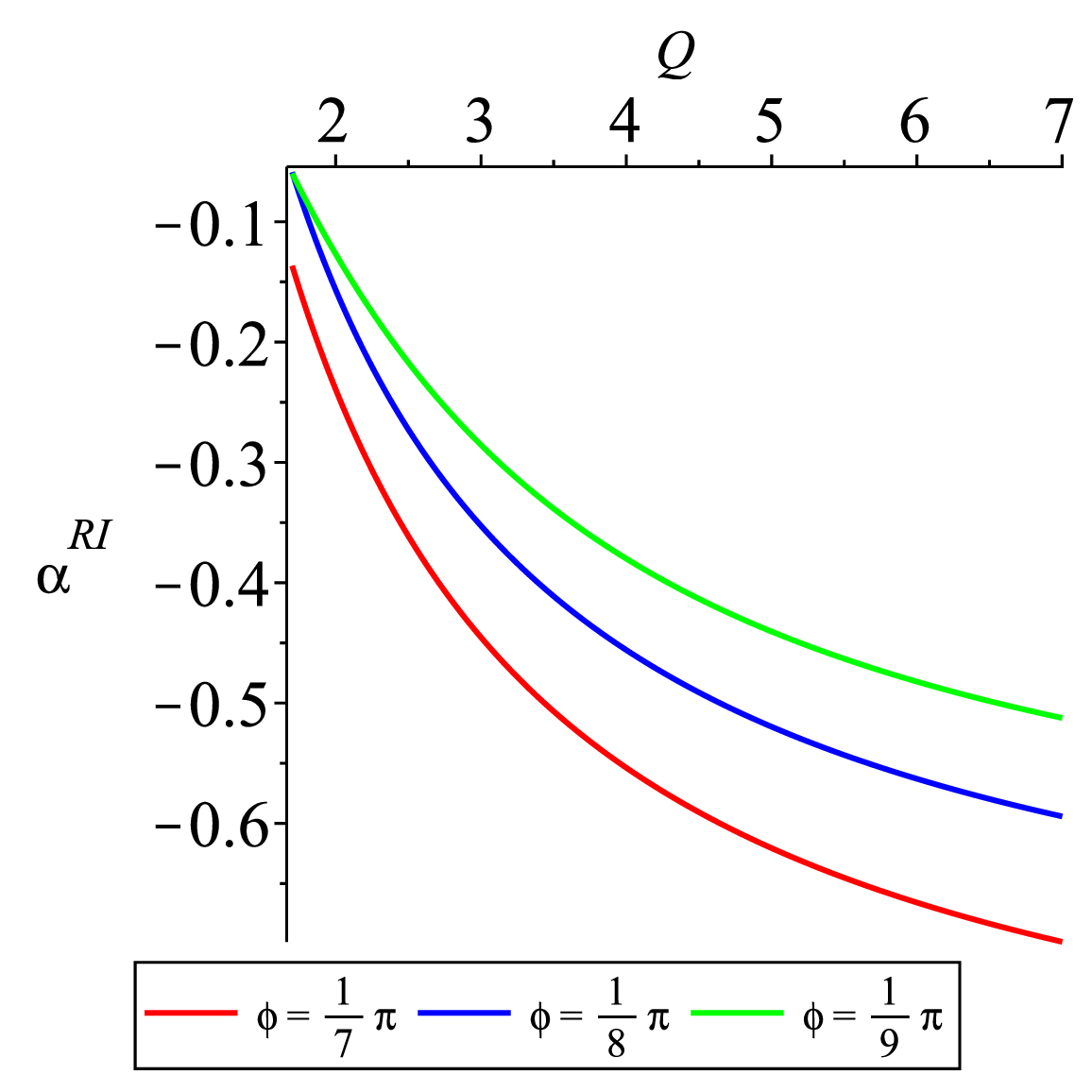}
\caption{Deflection angle $\alpha^{RI}$ as a function of $b$, $v$, $m$ (top row, left to right) and $r_s$, $q$, $Q$ (middle and bottom rows, left to right) for $q>0$.}\label{m4}
\end{figure}

Equation~(\ref{geo}), and the equations derived from it below, are written with $A_t$ retained as an explicit symbol rather than being substituted immediately by $A_t=-Q/r$. This is a deliberate choice of presentation: Eq.~(\ref{geo}) and its derivatives are general results, valid for any static, spherically symmetric electrostatic potential $A_t(r)$ sourced by the metric functions $A(r)$, $B(r)$, and $C(r)$ of Eq.~(2). Keeping $A_t$ symbolic at this stage keeps the structure of the trajectory equation transparent and makes explicit which terms originate from the electromagnetic coupling (through $A_t$) as opposed to the purely gravitational terms (through $A$, $B$, $C$). Only once the specific charged-wormhole background of Eq.~(\ref{e1}) is adopted do we substitute the explicit form $A_t=-Q/r$; this substitution is carried out numerically when generating Fig.~\ref{m4}, and $U_0$ in Eq.~(5) is obtained by solving the resulting transcendental algebraic equation numerically for each choice of the physical parameters, since a closed-form solution for $U_0$ is not available.

Differentiating Eq.~(\ref{geo}) with respect to $\phi$ gives
\begin{widetext}
\begin{equation}
\frac{d^2 U}{d\phi^2}+U=\frac{\left(1+\frac{\sqrt{1-v^2}q A_{t}}{m}\right)^2 -QU(1-v^2+v^2b^2U^2)^4-1}{v^2b^2(Q^2U^2+1)\!\left(1+Q^2U^2-U\!\left(-8r_{s}\!\left(\frac{1}{U^2}+\frac{2r_{s}}{U}+2r^{2}_{s}\right)e^{-\frac{1}{Ur_{s}}}\pi \rho_{s}+C_1\right)\right)}.
\end{equation}
\end{widetext}

Since planetary orbits are nearly circular, we first consider the unperturbed circular orbit, for which the reciprocal radial coordinate is constant, $U(\phi)=U_0$. Hence,
$
\frac{dU}{d\phi}=0,\
\frac{d^2U}{d\phi^2}=0,
$
and substituting these conditions into Eq.~(4) yields

\begin{widetext}
\begin{equation}
    U_0=\frac{\left(1+\frac{\sqrt{1-v^2}q A_{t}}{m}\right)^2 -QU_{0}(1-v^2+v^2b^2U_{0}^2)^4-1}{v^2b^2(Q^2U_{0}^2+1)\!\left(1+Q^2U_{0}^2-U_{0}\!\left(-8r_{s}\!\left(\frac{1}{U_{0}^2}+\frac{2r_{s}}{U_{0}}+2r^{2}_{s}\right)e^{-\frac{1}{U_{0}r_{s}}}\pi \rho_{s}+C_1\right)\right)}.
\end{equation}
\end{widetext}

With $A_t=-Q/r=-QU_0$ substituted explicitly, Eq.~(5) becomes a transcendental algebraic equation for $U_0$ that admits no closed-form inversion. For each choice of the physical parameters $(b,v,m,q,Q,r_s,\rho_s)$ we therefore solve Eq.~(5) for $U_0$ numerically; the resulting value is used as the background circular-orbit radius about which the perturbative expansion below is carried out.

To investigate small deviations from the circular orbit and determine the perihelion shift, we introduce a small perturbation about the unperturbed solution by writing
$
U=U_0(1+\epsilon),\qquad |\epsilon|\ll1,
$
where $\epsilon$ is the perturbation parameter. Substituting this into Eq.~(4) and retaining only terms up to first order in $\epsilon$, we obtain

\begin{widetext}
\begin{equation}
 \frac{d^2 \epsilon}{d\phi^2}+\epsilon= \Bigg[\frac{\left(1+\frac{\sqrt{1-v^2}q A_{t}}{m}\right)^2 -QU_{0}(1-v^2+v^2b^2U_{0}^2)^4-1}{U_0v^2b^2(Q^2U_{0}^2+1)\!\left(1+Q^2U_{0}^2-U_{0}\!\left(-8r_{s}\!\left(\frac{1}{U_{0}^2}+\frac{2r_{s}}{U_{0}}+2r^{2}_{s}\right)e^{-\frac{1}{U_{0}r_{s}}}\pi \rho_{s}+C_1\right)\right)}-1\Bigg].
\end{equation}  
\end{widetext}

\begin{widetext}
The perturbative solution for the trajectory is
\begin{multline}
  U=U_0\Biggr[1+C_2\cos\phi+C_3\sin\phi +\frac{1}{m^2v^2b^2(1+Q^2U^2_{0})\!\left(\!\left(U_0-C_1U_0^2+Q^2U^3_0\right)e^{\frac{1}{U_0r_s}}+16\rho_s \pi r_s\!\left(\frac{1}{2}+U_0r_s+U^2_0r^2_s\right)\!\right)} \\
   \times \Biggr\{2A_tmqe^{\frac{1}{U_0r_s}}\sqrt{1-v^2} -QU_0m^2\!\left(b^2U^2_0-1\right)v^2)^4e^{\frac{1}{U_0r_s}}
   +\bigg(-Q^4b^2m^2v^2U^5_0+Q^2C_1b^2m^2v^2U^4_0-2Q^2b^2m^2v^2U^3_0 \\+C_1b^2m^2v^2U^2_0-b^2m^2v^2U_0-A^2_tq^2v^2+A^2_tq^2\bigg)e^{\frac{1}{U_0r_s}}
   -16\rho_sm^2v^2b^2r_s\pi(1+Q^2U^2_0)\!\left(\frac{1}{2}+r_sU_0+r^2_sU^2_0\right)
   \Biggr\} \Biggr].
\end{multline}
\end{widetext}

As in Eqs.~(4)--(6), $A_t$ appears here symbolically, with $A_t=-QU_0$ substituted at the point of numerical evaluation.

\begin{figure}[htbp]
\centering
\includegraphics[width=0.85\columnwidth]{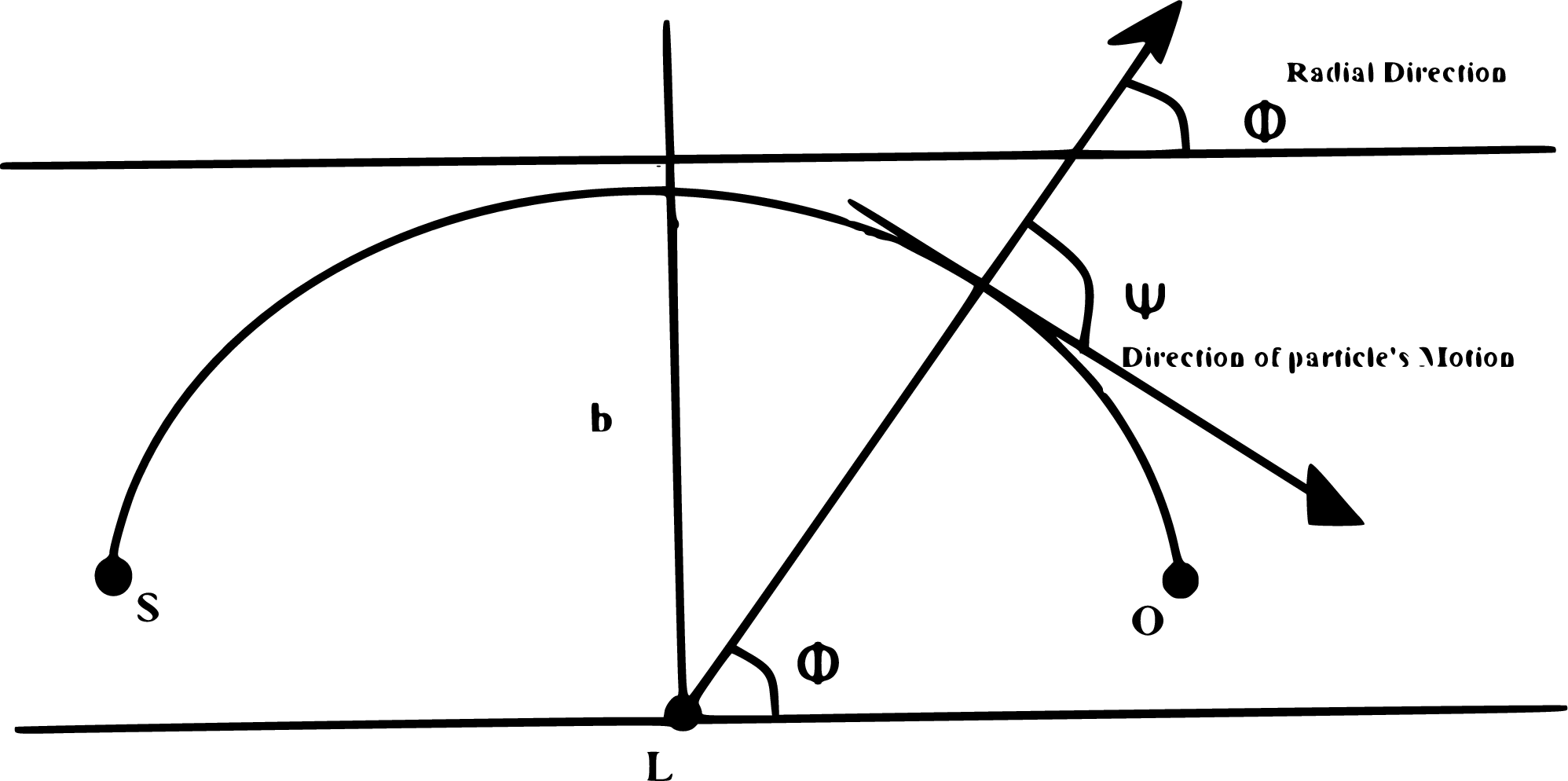}
\caption{Orbital motion of a charged massive particle around the charged galactic wormhole.}\label{c}
\end{figure}

In the Rindler-Ishak method, we determine the angle $\psi$ between the radial direction ($x^i$) and the direction of particle motion ($y^i$):
\begin{equation}
\cos\psi=\frac{g_{ij}x^iy^j}{\sqrt{g_{ij}x^ix^j}\sqrt{g_{ij}y^iy^j}},\label{eq36}
\end{equation}
with $t=\text{const}$, $\theta=\pi/2$. Setting $x^i=(\gamma,1)d\phi$ and $y^i=(1,0)dr$ where $\gamma=dr/d\phi$:
\begin{equation}
  \tan\psi =\frac{\sqrt{g_{\phi\phi}/g_{rr}}}{|\gamma|}.
\end{equation}
As shown in Fig.~\ref{c}, the one-sided deflection angle is given by $\alpha=\psi(\phi)-\phi$, and the total deflection is $2\alpha$. The Rindler-Ishak deflection angle is

\begin{equation}\label{m1}
\alpha^{RI}=2\arctan\!\left(\frac{\left(\frac{g_{\phi\phi}}{g_{rr}}\right)^{\frac{1}{2}}}{\left|-{U^{-2}(\phi)}\,\frac{dU(\phi)}{d\phi}\right|}\right)-2\phi.
\end{equation}

Figure~\ref{m4} shows the dependence of the deflection angle on the impact parameter $b$, particle velocity $v$, particle mass $m$, scale radius $r_s$, central density $\rho_s$, and wormhole charge $Q$ for a positively charged particle. The magnitude of the deflection decreases with increasing $b$ and $v$, since particles passing farther from the wormhole or moving at higher velocities experience weaker effective gravitational and electromagnetic interactions. In contrast, increasing the particle mass, central density, or wormhole charge enhances the deflection owing to the stronger gravitational and electromagnetic fields. For a negatively charged particle ($q<0$), the same six dependencies are mirrored in sign: the deflection angle decreases with increasing $b$, $v$, and $m$, and increases with $r_s$, $|q|$, and $Q$. This sign reversal follows directly from the change of sign of the electromagnetic coupling term $qA_t$ in Eq.~(\ref{geo}), while the underlying functional form of each dependence is otherwise unchanged. Although the qualitative behavior remains the same for both charge signs, reversing the particle charge modifies the electromagnetic coupling with the charged wormhole, resulting in quantitatively different deflection angles.

\section{CMP Deflection via the Gauss-Bonnet Method} \label{3}
The Gauss-Bonnet method is particularly interesting since the deflection of the particle is independent of coordinates. Using the Jacobi metric $J_{ij}$ developed from the Riemannian spacetime, one can define an area on the Jacobi reference geometry and find the deflection. The Gaussian curvature is \cite{r56}:
\begin{equation}\label{37}
    \begin{split}
K =&- \frac{1}{\sqrt{J_{rr}J_{\phi\phi}}}\!\left[\frac{\partial}{\partial r}\!\left(\frac{1}{\sqrt{J_{rr}}}\frac{\partial\sqrt{J_{\phi\phi}}}{\partial r}\right)\right]\\
 - & \frac{1}{\sqrt{J_{rr}J_{\phi\phi}}}\!\left[\frac{\partial}{\partial\phi}\!\left(\frac{1}{\sqrt{J_{\phi\phi}}}\frac{\partial\sqrt{J_{rr}}}{\partial\phi}\right)\right].
\end{split}
\end{equation}

\begin{figure}[htbp]
\centering
\includegraphics[scale=0.19]{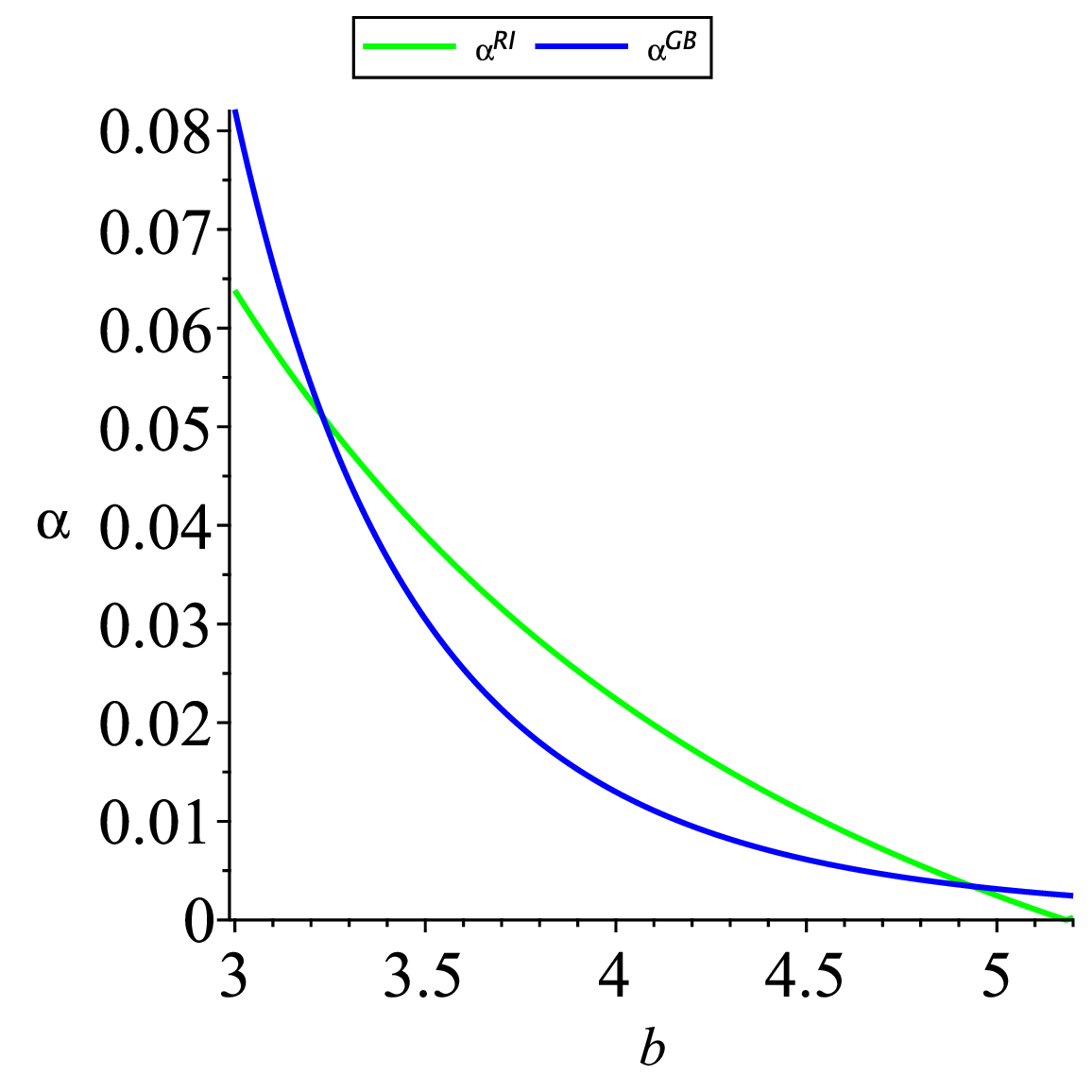}
\includegraphics[scale=0.19]{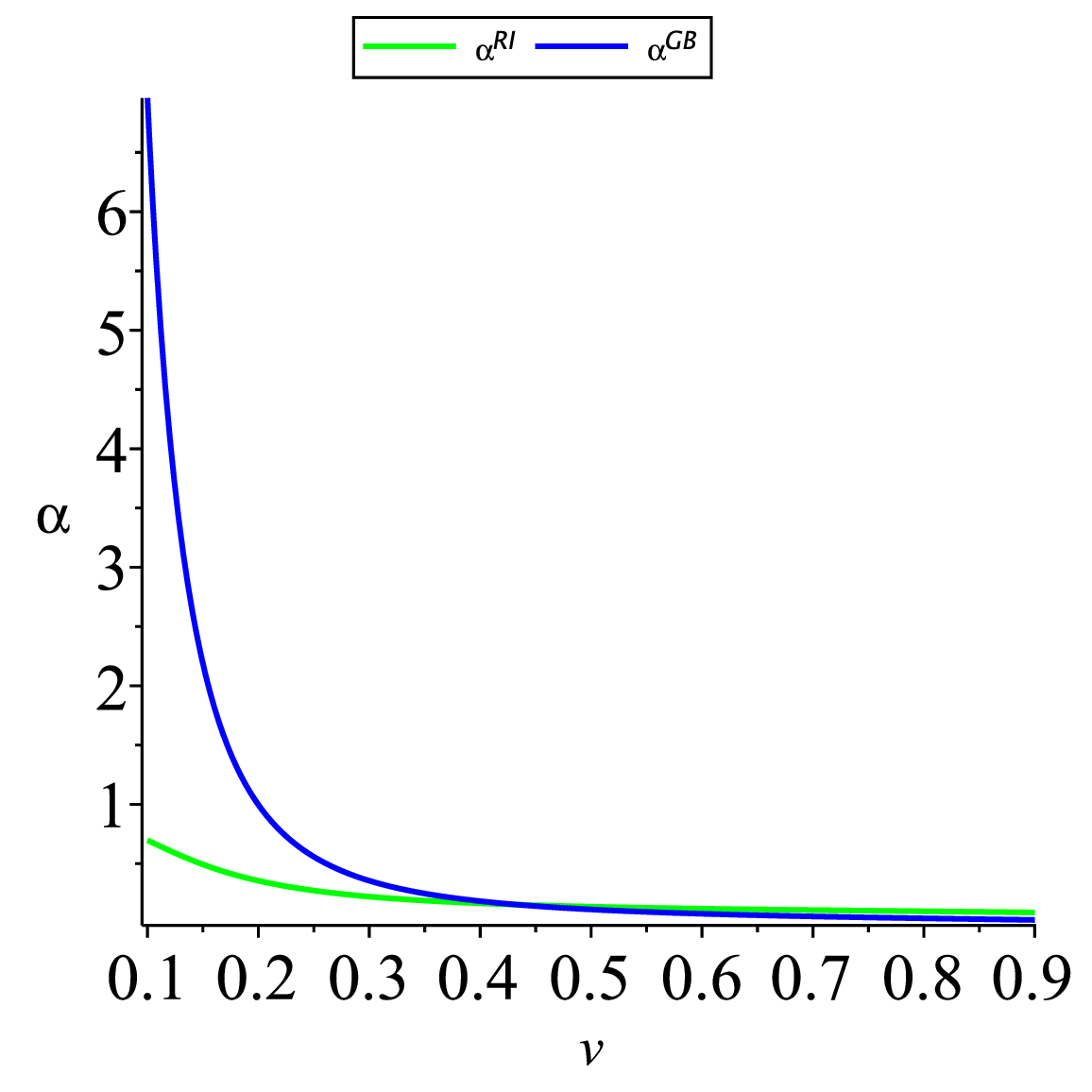}\\[-2pt]
\includegraphics[scale=0.19]{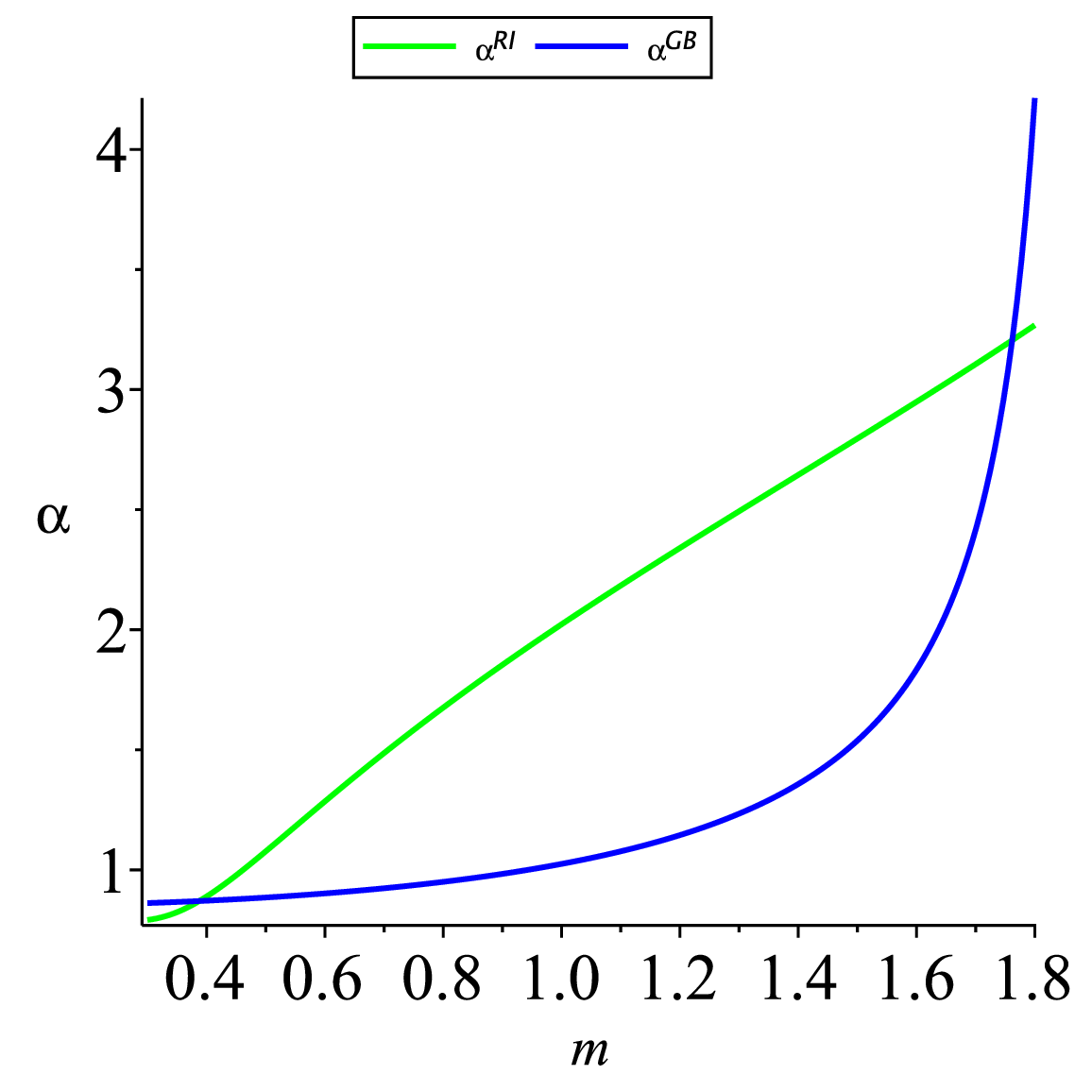}
\includegraphics[scale=0.19]{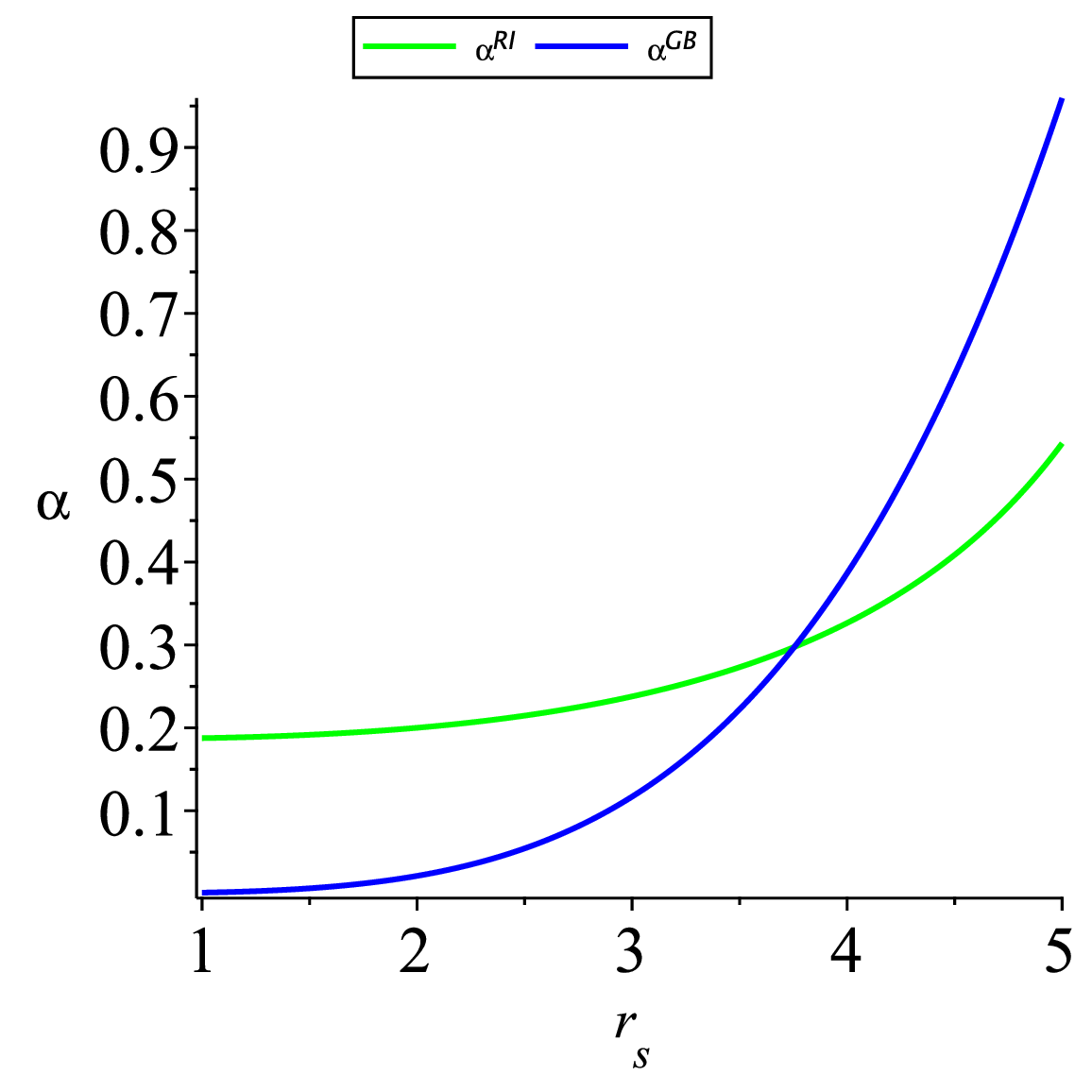}\\[-2pt]
\includegraphics[scale=0.19]{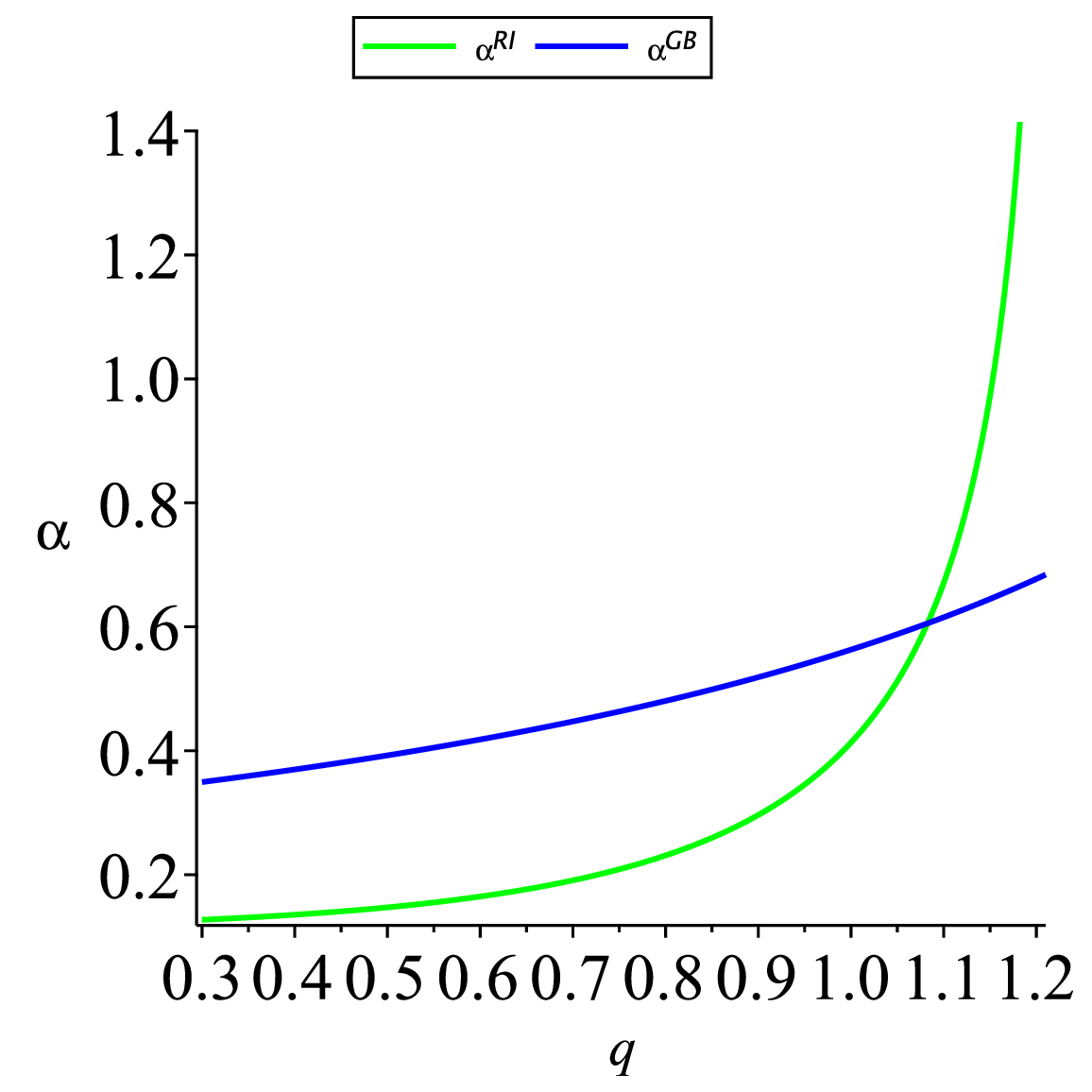}
\includegraphics[scale=0.19]{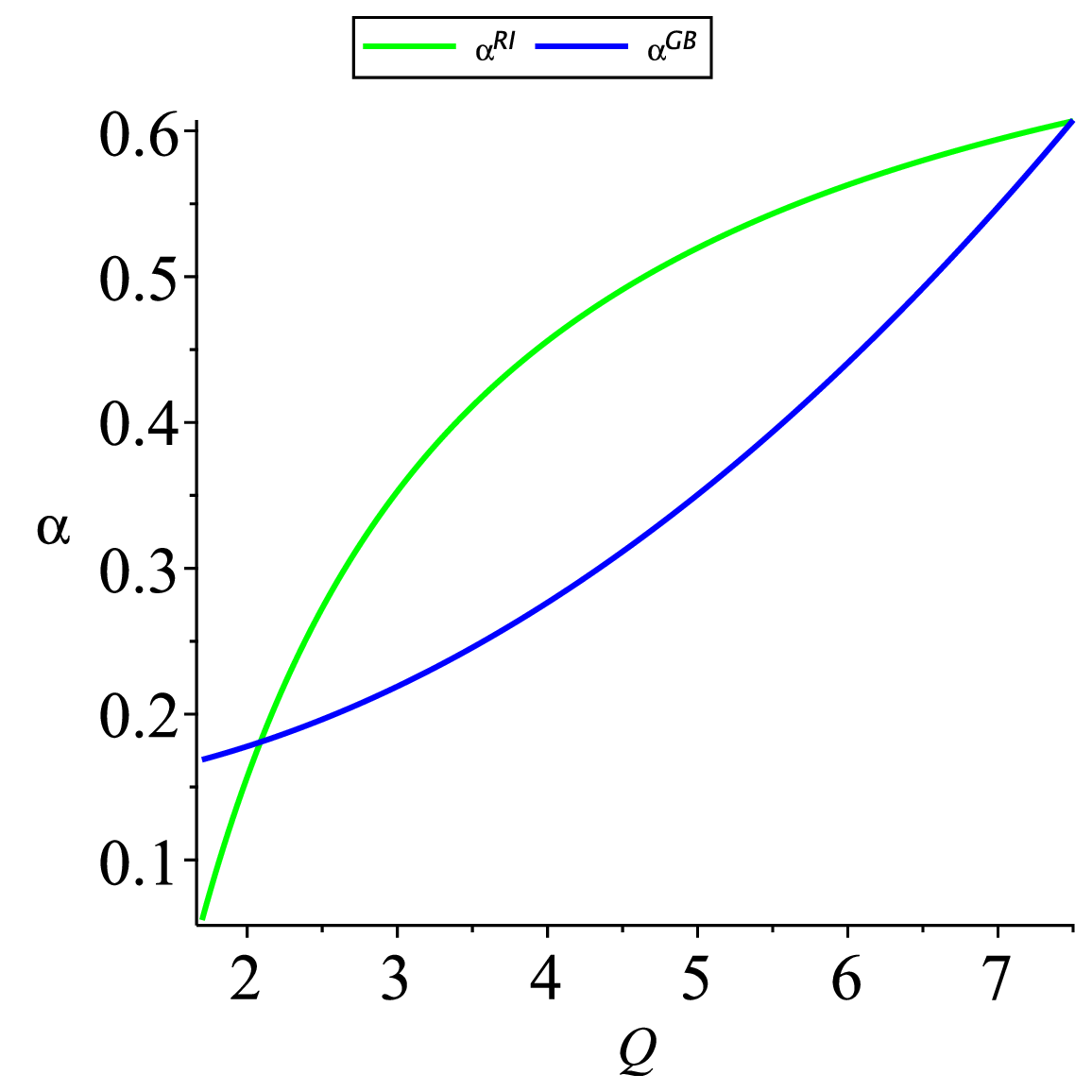}
\caption{Comparison of the deflection angles $\alpha^{RI}$ and $\alpha^{GB}$ as functions of $b$, $v$, $m$ (top row, left to right) and $r_s$, $q$, $Q$ (middle and bottom rows, left to right) for the representative case $q=-0.6$.}\label{m8}
\end{figure}

The deflection angle is determined by evaluating $K$ over the Riemannian-Jacobi geometry (asymptotically Euclidean):
\begin{equation}\label{38}
\alpha =    - \int_{0}^{\pi} \int_{r_0}^{\infty} K \sqrt{\det J} \,dr\,d\phi.
\end{equation}
Using the perturbation approach with $r_0=b/\sin\phi$ as the zeroth-order trajectory:
\begin{equation}\label{m2}
\alpha^{GB} \approx    - \int_{0}^{\phi} \int_{{b}/{\sin\phi}}^{\infty} K \sqrt{\det J} \,dr\,d\phi.
\end{equation}
Due to the complicated form of this integral, we discuss the nature of $\alpha^{GB}$ graphically together with $\alpha^{RI}$ in Fig.~\ref{m8}. Similar to the Rindler--Ishak method, the magnitude of the deflection decreases with increasing impact parameter and particle velocity, while it increases with the particle mass, central density, and wormhole charge, due to the enhanced gravitational and electromagnetic interactions. For $q>0$ a surplus deflection angle is observed, increasing with $b$, $v$, and $m$, and decreasing with $r_s$, $q$, and $Q$; for $q<0$ the opposite monotonic trends are found, mirroring the pattern already established for $\alpha^{RI}$ in Sec.~\ref{2}. Although the qualitative behavior is identical for both signs of the particle charge, reversing the charge modifies the electromagnetic coupling with the charged wormhole, leading to quantitatively different deflection angles. The excellent qualitative agreement between the Gauss--Bonnet and Rindler--Ishak methods demonstrates the consistency and robustness of our results.

\section{Photon Ring of charged galactic wormhole} \label{4}

The celestial coordinates \cite{Chandrasekhar1983} are defined as
\begin{equation}
\alpha = \lim_{r_{0}\to\infty}\!\left(-r_{0}^{2}\sin\theta_{0}\,\frac{d\phi}{dr}\right) = -\xi\csc\theta_{0},
\end{equation}
\begin{equation}
\beta = \lim_{r_{0}\to\infty}\!\left(r_{0}^{2}\frac{d\theta}{dr}\right) = \sqrt{\eta-\xi^{2}\csc^{2}\theta_{0}}.
\end{equation}
The photon ring $\mathcal{S}$ is
\begin{equation}
\mathcal{S}(\alpha,\beta) := \alpha^{2}+\beta^{2} = \eta = \frac{r^{2}}{1+{Q^{2}}/{r^{2}}}.
\label{eq:37}
\end{equation}
This demonstrates that the shadow radius is exclusively governed by the wormhole charge parameter $Q$. As $Q$ increases, the shadow radius decreases monotonically, indicating that larger charge values intensify the gravitational field, enhancing inward deflection of photon trajectories.

\begin{figure}[htbp]
    \centering
    \begin{subfigure}{0.44\columnwidth}
        \includegraphics[width=\linewidth]{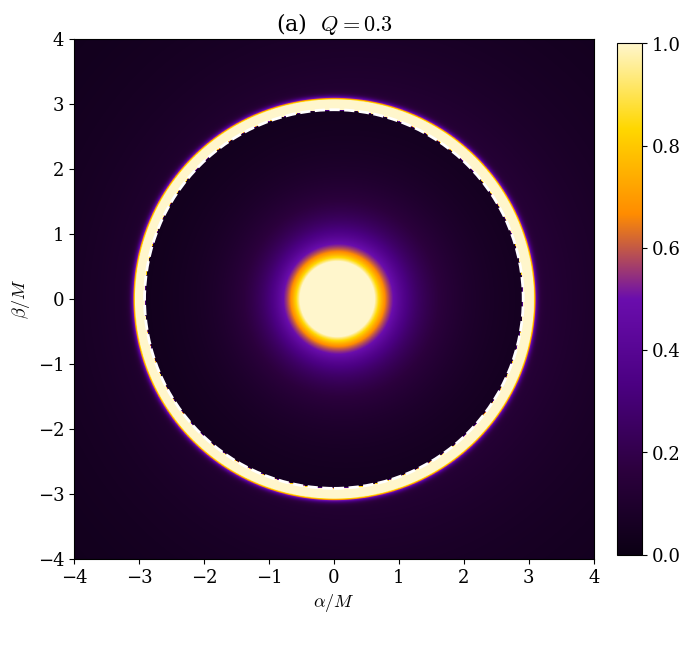}
        \caption{$Q=0.3$}
    \end{subfigure}
    \hfill
    \begin{subfigure}{0.44\columnwidth}
        \includegraphics[width=\linewidth]{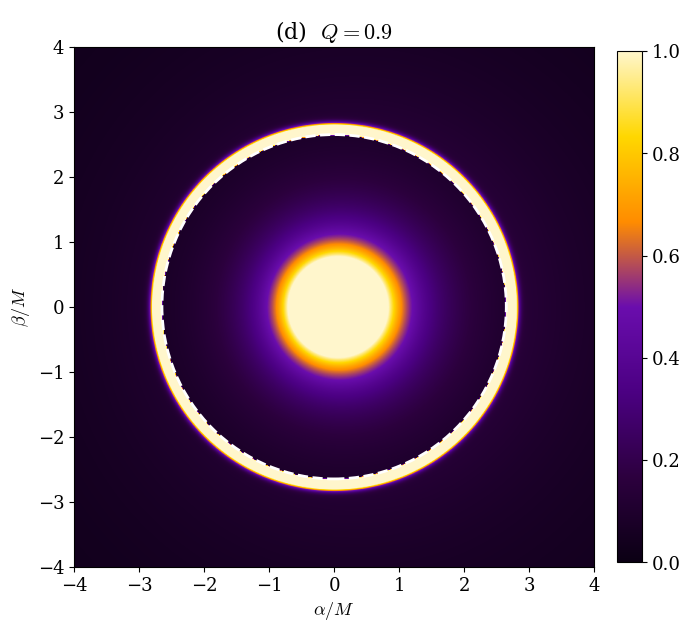}
        \caption{$Q=0.9$}
    \end{subfigure}
    \\[2pt]
    \begin{subfigure}{0.44\columnwidth}
        \includegraphics[width=\linewidth]{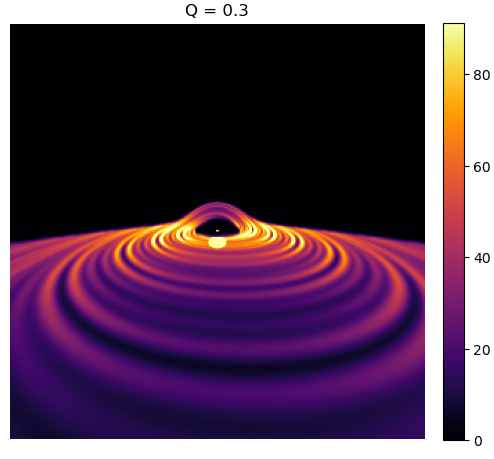}
        \caption{$Q=0.3$}
    \end{subfigure}
    \hfill
    \begin{subfigure}{0.44\columnwidth}
        \includegraphics[width=\linewidth]{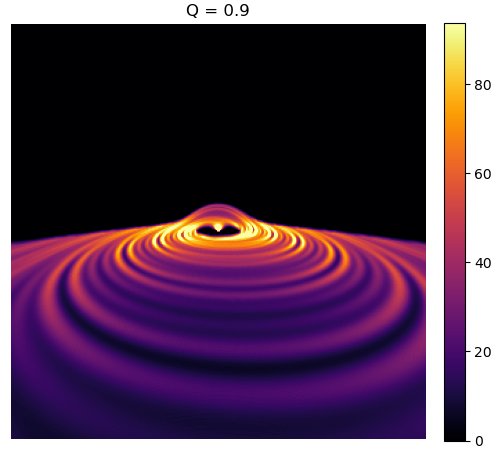}
        \caption{$Q=0.9$}
    \end{subfigure}
    \caption{Top row: shadow and photon-ring structure of the optically thin accretion flow. Bottom row: corresponding fully ray-traced images of the same accretion flow. Both rows are shown for $Q=0.3$ (left) and $Q=0.9$ (right).}
    \label{fig:combined_rt}
\end{figure}

The ray-tracing analysis presented in Fig.~\ref{fig:combined_rt} was carried out using a backward (image-plane-to-source) relativistic ray-tracing scheme. For each pixel in the observer's image plane, characterized by celestial coordinates $(\alpha,\beta)$, a past-directed null geodesic was integrated numerically from the observer at $r_0\to\infty$ toward the wormhole using the geodesic equations derived from the metric in Eq.~(\ref{e1}), with the impact parameters fixed by the initial conditions $\alpha=-\xi\csc\theta_0$ and $\beta=\sqrt{\eta-\xi^2\csc^2\theta_0}$. The accretion flow surrounding the wormhole was modeled as an optically and geometrically thin equatorial disk, following the standard treatment adopted for horizonless and black-hole compact objects, with a radial emissivity profile that peaks near the photon sphere/throat region and falls off at larger radii. Each time a geodesic crossed the equatorial plane ($\theta=\pi/2$), the corresponding emitted specific intensity was evaluated using the local emission profile and accumulated, weighted by the appropriate redshift factor obtained from the metric coefficients $g_{tt}$ and $g_{\phi\phi}$ along the photon trajectory, thereby incorporating gravitational redshift and Doppler boosting self-consistently. Multiple crossings of the equatorial plane were included to capture both the direct image and the higher-order (photon-ring) images produced by photons that wind around the wormhole before reaching the observer. The resulting intensity map was binned onto the observer's image plane to produce the ray-traced images in Fig.~\ref{fig:combined_rt}, which presents both the shadow/photon-ring structure and the corresponding fully ray-traced accretion image together, for $Q=0.3$ and $Q=0.9$. Because the underlying wormhole spacetime is static and spherically symmetric, the shadow boundary itself is exactly circular for all values of $Q$, as follows analytically from Eq.~(\ref{eq:37}); the qualitative interest in this section therefore lies not in the shadow shape, which is trivially circular in the absence of rotation, but in the $Q$-dependence of the shadow radius, the photon-ring radius, and the intensity/redshift structure of the ray-traced accretion image, all of which vary non-trivially with the wormhole charge and thus encode the charge parameter observationally even though the outer silhouette remains circular.

Figure~\ref{fig:combined_rt} (top row) illustrates the photon-ring and shadow structures for two values of the electric charge. As the charge increases from $Q=0.3$ to $Q=0.9$, a clear contraction of the photon ring is observed. The shadow region becomes progressively more compact, indicating that the presence of charge enhances the effective spacetime curvature experienced by photons. Furthermore, the intensity contrast between the central bright region and the surrounding photon ring becomes more pronounced for larger values of $Q$, demonstrating that the electromagnetic contribution plays a significant role in shaping the observable image. Figure~\ref{fig:combined_rt} (bottom row) presents the fully ray-traced images of the optically thin accretion flow. As $Q$ increases, the bright ring becomes sharper and more compressed toward the central region, while the outer emission layers exhibit stronger warping effects due to enhanced deflection of photon trajectories. The progressive modification of the emission profile with increasing $Q$ highlights the interplay between gravitational and electromagnetic fields, and suggests that such charge-dependent distortions could serve as important diagnostics in distinguishing charged wormholes from other compact astrophysical objects.

\section{Conclusion} \label{5}

In this work, we have investigated the deflection of charged massive particles propagating in the spacetime of a charged galactic wormhole by employing two complementary approaches, namely the Rindler--Ishak method and the Gauss--Bonnet theorem.

\begin{itemize}
    \item Within the Rindler--Ishak framework, the deflection angle was derived in Eq.~\eqref{m1}. As illustrated in Fig.~\ref{m4}, for $q>0$ a surplus deflection angle $\alpha^{\mathrm{RI}}$ signals effective repulsive behavior: the deflection angle increases with the impact parameter $b$, particle velocity $v$, and mass parameter $m$, while decreasing with the particle charge $q$, wormhole charge $Q$, and scale radius $r_s$. For $q<0$, the deflection angle exhibits conventional behavior, decreasing with $b$, $v$, and $m$, and increasing with $r_s$, $q$, and $Q$.

    \item The Gauss--Bonnet deflection angle given in Eq.~\eqref{m2} follows the same qualitative parameter dependence as $\alpha^{RI}$ for both signs of $q$, as discussed in Sec.~\ref{3}.

    \item Fig.~\ref{m8} presents a direct comparison between the Gauss--Bonnet and Rindler--Ishak methods. For the representative case $q=-0.6$, both methods exhibit excellent qualitative agreement in the behavior of the deflection angle. The Rindler--Ishak method generally predicts slightly larger deflection angles for variations in $b$, $m$, $r_s$, and $Q$, while yielding comparatively smaller values for varying particle charge.
An important outcome of this work is that the deflection angles predicted by the Rindler--Ishak and Gauss--Bonnet methods become nearly indistinguishable in the relativistic regime. This convergence is not merely a numerical observation but follows directly from the analytical structure of the trajectory equation: the velocity-dependent correction terms proportional to $(1-v^2)$ are progressively suppressed as $v\rightarrow1$, causing both methods to recover the same null-geodesic (photon) limit. In contrast, the finite contribution of these velocity-dependent corrections in the low-velocity regime naturally accounts for the observed differences between the two approaches, demonstrating that they arise from the underlying dynamics rather than from any inconsistency between the methods.

    \item As illustrated in Fig.~\ref{fig:combined_rt} (top row), the ray-traced shadow and photon-ring morphology exhibit a pronounced dependence on the wormhole charge parameter $Q$. As $Q$ increases, the photon ring becomes progressively more compact and sharply defined, reflecting stronger gravitational lensing and enhanced spacetime curvature in the vicinity of the wormhole. The corresponding intensity distribution indicates more efficient photon trapping near the wormhole throat, leading to noticeable modifications in the shadow structure. These features suggest that shadow morphology may serve as a valuable observational probe for distinguishing charged wormholes from other compact astrophysical objects.

    \item As illustrated in Fig.~\ref{fig:combined_rt} (bottom row), the ray-tracing analysis of optically thin accretion flows reveals that the wormhole charge parameter $Q$ plays a significant role in shaping the observed intensity distribution and ring-like emission structure. As $Q$ increases, the lensed image becomes progressively more compact and pronounced, reflecting enhanced gravitational lensing and stronger spacetime curvature in the vicinity of the wormhole.

    \item Overall, while the Rindler--Ishak and Gauss--Bonnet approaches differ in their underlying geometric formulations, they provide consistent and complementary descriptions of charged particle deflection, reinforcing the reliability of the obtained results.
\end{itemize}

Our study opens several promising directions for future research. Astrophysical compact objects are generally expected to be nearly electrically neutral in realistic astrophysical environments, since any significant net charge would be rapidly neutralized by ambient plasma. The charged wormhole studied here should therefore be understood as an idealized theoretical construct rather than a claim about realistic astrophysical wormholes. Its value lies in the same sense that the Reissner--Nordstr\"om black hole is valuable in black-hole physics: as an exactly solvable, controlled extension of the neutral case that isolates the effect of a single additional parameter, the charge $Q$, on the geometry and on observable quantities such as the deflection angle, shadow radius, and photon-ring structure. This controlled setting allows us to identify, in isolation, which observational signatures --- for example, the $Q$-dependent contraction of the photon ring and the associated changes in the deflection angle reported above --- are attributable specifically to the electromagnetic sector of the theory, rather than to the gravitational sector alone. Such idealized solutions are standard practice in the compact-object literature and serve as necessary benchmarks before more realistic, plasma-screened, or magnetically supported charge configurations can be modeled. We adopt this same rationale here, and view the present charged-wormhole analysis as a stepping stone toward more astrophysically realistic extensions, several of which we outline below. A more comprehensive investigation of charged particle dynamics and radiative processes may uncover characteristic signatures that distinguish charged wormholes from black holes and other compact objects. Furthermore, extending the present analysis to rotating charged wormholes, modified theories of gravity, plasma environments, and more realistic galactic matter distributions would further enhance our understanding of these exotic geometries and their observational consequences. As the precision of astronomical observations continues to improve, theoretical predictions of lensing, shadow, photon-ring, and accretion signatures derived from such models will provide valuable benchmarks for interpreting future observations and for assessing the viability of charged wormholes as theoretical solutions of the Einstein--Maxwell equations.

\section*{Acknowledgments}
FR , MKH and AS would like to thank the authorities of the Inter-University Centre for Astronomy and Astrophysics, Pune, India for providing the research facilities. FR also gratefully acknowledges support by RUSA2.0 and ANRF-SERB, Govt.\ of India.


\begin{thebibliography}{}
\bibitem{r57} Y.~Sofue, Publ.\ Astron.\ Soc.\ Jpn.\ \textbf{65}, 118 (2013).

\bibitem{b1}
Md~K.~Hossain and F.~Rahaman,
Int.\ J.\ Geom.\ Meth.\ Mod.\ Phys.\ \textbf{23}, 2550151 (2026).


\bibitem{r1} A.~Einstein, Science \textbf{84}, 506 (1936).

\bibitem{r2} P.~Schneider, J.~Ehlers, and E.~E.~Falco, \textit{Gravitational Lenses} (Springer-Verlag, New York, 1992).

\bibitem{r3} R.~W.~Fuller and J.~A.~Wheeler, Phys.\ Rev.\ \textbf{128}, 919 (1962).

\bibitem{r4} A.~Einstein and N.~Rosen, Phys.\ Rev.\ \textbf{48}, 73 (1935).

\bibitem{r5} M.~S.~Morris and K.~S.~Thorne, Am.\ J.\ Phys.\ \textbf{56}, 395 (1988).

\bibitem{r6} J.~A.~Gonzalez, F.~S.~Guzman, and O.~Sarbach, Class.\ Quantum Grav.\ \textbf{26}, 015011 (2008).

\bibitem{r7} T.~Ohgami and N.~Sakai, Phys.\ Rev.\ D \textbf{91}, 124020 (2015).

\bibitem{r8} M.~Visser, \textit{Lorentzian Wormholes: From Einstein to Hawking} (Woodbury, 1995).

\bibitem{r9} K.~S.~Virbhadra and G.~F.~R.~Ellis, Phys.\ Rev.\ D \textbf{62}, 084003 (2000).

\bibitem{r10} V.~Bozza et al., Gen.\ Rel.\ Grav.\ \textbf{33}, 1535 (2001).

\bibitem{r11} V.~Bozza, Phys.\ Rev.\ D \textbf{66}, 103001 (2002).

\bibitem{r12} V.~Bozza, Gen.\ Rel.\ Grav.\ \textbf{42}, 2269 (2010).

\bibitem{r13} N.~Tsukamoto, Phys.\ Rev.\ D \textbf{95}, 064035 (2017).

\bibitem{r14} A.~Abdujabbarov et al., Int.\ J.\ Mod.\ Phys.\ D \textbf{26}, 1741011 (2017).

\bibitem{r15} H.~Chakrabarty et al., Phys.\ Rev.\ D \textbf{98}, 024022 (2018).

\bibitem{r16} C.~A.~Benavides-Gallego, A.~A.~Abdujabbarov, and C.~Bambi, Eur.\ Phys.\ J.\ C \textbf{78}, 1 (2018).

\bibitem{r17} B.~Bhawal and S.~Kar, Phys.\ Rev.\ D \textbf{46}, 2464 (1992).

\bibitem{r18} F.~Rahaman, M.~Kalam, and A.~Ghosh, arXiv:gr-qc/0605095 (2006).

\bibitem{r19} H.~Maeda and M.~Nozawa, Phys.\ Rev.\ D \textbf{78}, 024005 (2008).

\bibitem{r20} F.~S.~N.~Lobo and M.~A.~Oliveira, Phys.\ Rev.\ D \textbf{80}, 104012 (2009).

\bibitem{r21} M.~H.~Dehghani and S.~H.~Hendi, Gen.\ Rel.\ Grav.\ \textbf{41}, 1853 (2009).

\bibitem{r22} P.~Kanti, B.~Kleihaus, and J.~Kunz, Phys.\ Rev.\ Lett.\ \textbf{107}, 271101 (2011).

\bibitem{r23} N.~M.~Garcia and F.~S.~N.~Lobo, Class.\ Quantum Grav.\ \textbf{28}, 085018 (2011).

\bibitem{r24} C.~G.~Boehmer, T.~Harko, and F.~S.~N.~Lobo, Phys.\ Rev.\ D \textbf{85}, 044033 (2012).

\bibitem{r25} T.~Harko et al., Phys.\ Rev.\ D \textbf{87}, 067504 (2013).

\bibitem{r26} S.~Kar, S.~Lahiri, and S.~SenGupta, Phys.\ Lett.\ B \textbf{750}, 319 (2015).

\bibitem{r27} K.~A.~Bronnikov and A.~M.~Galiakhmetov, Grav.\ Cosm.\ \textbf{21}, 283 (2015).

\bibitem{r28} R.~Shaikh, Phys.\ Rev.\ D \textbf{92}, 024015 (2015).

\bibitem{r29} R.~Shaikh, Natl.\ Conf.\ CICAHEP, Dibrugarh, India (2015).

\bibitem{r30} C.~Bambi et al., Phys.\ Rev.\ D \textbf{93}, 064016 (2016).

\bibitem{r31} R.~Shaikh and S.~Kar, Phys.\ Rev.\ D \textbf{94}, 024011 (2016).

\bibitem{r32} R.~Myrzakulov et al., Class.\ Quantum Grav.\ \textbf{33}, 125005 (2016).

\bibitem{r33} P.~H.~R.~S.~Moraes, R.~A.~C.~Correa, and R.~V.~Lobato, JCAP \textbf{2017}, 029 (2017).

\bibitem{r34} H.~Moradpour, N.~Sadeghnezhad, and S.~H.~Hendi, Can.\ J.\ Phys.\ \textbf{95}, 1257 (2017).

\bibitem{r35} M.~Hohmann et al., JCAP \textbf{2018}, 003 (2018).

\bibitem{r36} P.~H.~R.~S.~Moraes and P.~K.~Sahoo, Phys.\ Rev.\ D \textbf{97}, 024007 (2018).

\bibitem{r37} R.~Shaikh, Phys.\ Rev.\ D \textbf{98}, 064033 (2018).

\bibitem{r38} A.~Jawad and H.~Moradpour, Int.\ J.\ Geom.\ Meth.\ Mod.\ Phys.\ \textbf{15}, 1850210 (2018).

\bibitem{r39} A.~Ovgun, K.~Jusufi, and I.~Sakalli, Phys.\ Rev.\ D \textbf{99}, 024042 (2019).

\bibitem{r40} M.~R.~Mehdizadeh and A.~H.~Ziaie, Phys.\ Rev.\ D \textbf{99}, 064033 (2019).

\bibitem{r41} M.~Zubair et al., Int.\ J.\ Geom.\ Meth.\ Mod.\ Phys.\ \textbf{16}, 1950046 (2019).

\bibitem{r42} N.~Godani and G.~C.~Samanta, Int.\ J.\ Mod.\ Phys.\ D \textbf{28}, 1950039 (2019).

\bibitem{r43} P.~H.~R.~S.~Moraes and P.~K.~Sahoo, Eur.\ Phys.\ J.\ C \textbf{79}, 1 (2019).

\bibitem{r44} M.~R.~Mehdizadeh and A.~H.~Ziaie, Mod.\ Phys.\ Lett.\ A \textbf{35}, 2050017 (2020).

\bibitem{r45} A.~Banerjee et al., arXiv:1904.10310 (2019).

\bibitem{r46} G.~Antoniou et al., arXiv:1904.13091 (2019).

\bibitem{r47} Event Horizon Telescope Collaboration, Astrophys.\ J.\ Lett.\ \textbf{875}, L1 (2019).

\bibitem{r48} K.~Akiyama, Astrophys.\ J.\ Lett.\ \textbf{875}, L2 (2019).

 



\bibitem{r52} J.~G.~Cramer et al., Phys.\ Rev.\ D \textbf{51}, 3117 (1995).

\bibitem{r51} M.~Safonova, D.~F.~Torres, and G.~E.~Romero, Phys.\ Rev.\ D \textbf{65}, 023001 (2001).

\bibitem{r53} J.~M.~Tejeiro and E.~Larranaga, arXiv:gr-qc/0505054 (2005).

\bibitem{r55} V.~Perlick, Living Rev.\ Relativ.\ \textbf{7}, 1 (2004).

\bibitem{r56} Md~K.~Hossain et al., Nucl.\ Phys.\ B (2024) 116598.

\bibitem{q1} F.~Rahaman et al., Eur.\ Phys.\ J.\ C \textbf{74}, 2750 (2014).

\bibitem{q2} F.~Rahaman et al., Ann.\ Phys.\ \textbf{350}, 561 (2014).





\bibitem{0} F.~Rahaman et al., J. Korean Phys. Soc. 86.12 (2025): 1204-1224.

\bibitem{Chandrasekhar1983}
S.~Chandrasekhar, \textit{The Mathematical Theory of Black Holes}, 1st ed.\ (Oxford University Press, Oxford, 1983).

\end{thebibliography}
\end{document}